# Engineering and Probing Atomic Quantum Defects in 2D Semiconductors – a perspective

Joshua A. Robinson[1-3] and Bruno Schuler[4,*]

[1] Department of Materials Science and Engineering, The Pennsylvania State University, University Park, PA 16802, USA

[2] Two-Dimensional Crystal Consortium, The Pennsylvania State University, University Park, PA 16802, USA

[3] Center for 2-Dimensional and Layered Materials, The Pennsylvania State University, University Park, PA 16802, USA

[4] nanotech@surfaces Laboratory, Empa – Swiss Federal Laboratories for Materials Science and Technology, 8600 Dübendorf, Switzerland

* bruno.schuler@empa.ch

## Abstract

Semiconducting two-dimensional (2D) transition metal dichalcogenides (TMDs) are considered a key materials class to scale microelectronics to the ultimate atomic level. The robust quantum properties in TMDs also enable new device concepts that promise to push quantum technologies beyond cryogenic environments. Mission-critical capabilities towards realizing these goals are the mitigation of accidental lattice imperfections and the deterministic generation of desirable defects. In this perspective, the authors review some of their recent results on engineering and probing atomic point defects in 2D TMDs. Furthermore we provide a personal outlook on the next frontiers in this fast evolving field.

## Introduction

Robust quantum phenomena in two-dimensional (2D) semiconductors are captivating an ever-growing number of scientists worldwide. These quantum states emerge from strong correlations imposed by the atomic confinement and reduced screening. Hence, 2D materials not only offer great promise to scale semiconductor technology to the ultimate atomic level, but also to explore new device paradigms based on quantum mechanical principles. Transition metal dichalcogenides (TMDs) in particular are a materials class that is unique in its facile synthesis, chemical diversity, inertness, and processability – qualities underpinning current efforts to integrate TMDs in next-generation semiconductor fabrication lines. Chemical and charge-transfer doping are the prime techniques to modulate the charge carrier concentration. Charge-transfer doping denotes the charge donation or withdrawal by an interfacing medium such as the substrate, dielectric overlayers, or adsorbed molecules. While virtually negligible for bulk semiconductors, this effect is often dominant for 2D semiconductors because of their all-surface nature and low density of states. For the same reasons electrostatic gating is highly effective at these reduced dimensions. In combination with atomically smooth interfaces formed between van der Waals (vdW) materials, electrostatic gating using vdW heterostructures is a non-invasive approach for uniformly tuning the global charge carrier concentration in TMDs[1].

Chemical doping, where foreign atoms are intentionally incorporated into the lattice, is indisputably the prime technology for doping bulk semiconductors. In ultrathin layers, however, defects (intentional or otherwise) remain largely unscreened such that their states are in general deeper and more tightly confined[2]. While this attribute limits the attainable free charge carrier concentration by chemical doping, it renders defect states in 2D semiconductors ideally suited as atomic quantum systems. The growth by chemical vapor deposition techniques unlock a versatile chemistry toolbox to allow designing such atomic quantum systems by chemical design principles. This is achieved by tweaking their spin, charge, lattice, and orbital degrees of freedom (see Figure 1).

Apart from zero dimensional point defects, some TMDs also exhibit a low formation energy of line defects in the form of mirror-twin boundaries. Such mirror-twin boundaries, detected for instance in $MoSe_2$ and $MoS_2$, host 1D conducting channels within the semiconductor band gap that can be described as a charge density wave[3] or Tomonaga-Luttinger liquid[4,5], respectively. More generally, TMDs can develop various types of structural[6,7] or compositional[8] domain boundaries that may host extended (topologically protected) conductive channels. Such well-defined 1D interfaces offer a vast engineering playground in itself that can drastically affect electronic device properties. In this perspective we focus the discussion on *point* defects.

Quantum properties of defects can be exploited in photonics and spintronics applications. Single point defects in TMDs are for instance extensively explored as single photon sources[9]. While early reports on single photon emission in TMDs are mainly attributed to a quantum dot-like confinement of excitons due to strain profiles[10–13], single photon emission due to point defects were recently reported[14]. This line of research connects to more advanced work on color centers in bulk insulator like diamond, SiC, and hBN, where one can read and manipulate the electron spin of certain defects by optical means[9,15]. 2D platforms offer benefits regarding placement control, photon extraction efficiency, tunability through external gates, and integration with photonic and plasmonic nanocavities. Moreover, spin-orbit split valence bands or moiré superlattices may be employed to mediate the non-local interaction between defects, acting as quantum bus. The strong spin-orbit coupling inherent to the heavy transition metals in tandem with spin-carrying defects are also attractive for magnetism and spintronics[16]. One example is gate-tunable ferromagnetism achieved by dilute vanadium doping of $WS_2$/$WSe_2$ at room temperature[17,18] that could be exploited in spin logic devices.

In all these applications a detailed knowledge of the defect atomic structure and electronic spectrum builds the foundation to understand defect functionality. Thereby, the atomic spatial and milli-electronvolt energy resolution of scanning probe microscopy is uniquely suited to probe surface defects in a non-invasive manner. In this perspective, we discuss current trends on the controlled generation and atomic-resolution characterization of defects in semiconducting TMDs based on the authors' own work and outline key challenges that lie ahead in this fast evolving field. We refer the reader to excellent topical reviews for a broader discussion on synthetic, optical, and device aspects of defect engineering in TMDs[16,19–22].

## Identifying Accidental Point Defects in As-Grown TMDs

Before any intentional doping can be targeted, a detailed understanding of accidental defects that are commonly present in as-synthesized samples is important. Using a combination of low-temperature scanning tunneling microscopy (STM) and spectroscopy (STS), noncontact atomic force microscopy (nc-AFM) with a carbon monoxide functionalized tip and *ab initio* density functional theory we were able to identify the most common defects that are consistently observed in CVD-grown $WS_2$ samples. In particular the discrete defect states in the band gap and their associated orbitals that can be mapped by STS, serve as a unique electronic fingerprint that can be compared to theoretical predictions. With CO-tip nc-AFM we can unambiguously assign the lattice site of the defect. The combination of these insights allowed us to assign the *a priori* unknown defect species as summarized in Figure 2. Specifically we find oxygen or CH substituting for sulfur ($O_S$[23], $CH_S$[24], respectively), as well as chromium and molybdenum substituting for tungsten ($Cr_W$[25], $Mo_W$[25]) in solid-source CVD-grown 2D layers. These are impurity defects with a very low formation energy. The STM topographies (at specific bias) of these defects are similar enough across a variety of TMDs, such that analog defects can be assign in MOCVD-grown $WSe_2$, MBE-grown $MoSe_2$ and exfoliated $MoS_2$ as well. While these are the most common defects in our samples, their abundance and relative frequency depends on the growth method and varies to a minor extent also between samples. Minimizing this "background" of accidental defects in synthetic TMDs is an important challenge that needs to be addressed.

A prominent absentee in the above list of common defects are chalcogen vacancies, which have often been proposed as the dominant defect species in literature. While chalcogen vacancies are the defect with the lowest formation energy (in absence of other reactants), and are indeed the first defect to form when annealing the sample in vacuum[26], this defect is very reactive and is likely to bind oxygen or hydrocarbons present during growth or in the atmosphere.

## Strategies for Intentional Point Defect Engineering

In Figure 3 we summarize different approaches we pursued to deliberately incorporate different types of defects with high specificity and density control, such as *in-vacuo* annealing[26], chemical doping[24,27], atom manipulation[28] and ion beam bombardment[29,30]. In particular we demonstrate substitutional doping with non-isovalent transition metals such as p-type V[31] and Nb[32], or n-type Re[27,33] and Mn[34] during synthesis. In the case of substitutional doping during growth of the 2D layer, the dopant tends to replace the transition metal (W or Mo) as it is similar in size. Importantly, however, the doping efficacy can be strongly impacted by the 2D/substrate interaction[34] and the dopant activation energy within the layer must also be considered for electronic applications. It is theoretically predicted[2] that the dopant ionization energy for p- and n-type dopants may be too large at the monolayer limit due to quantization effects where strong localization can limit ionization efficiency of the dopant, correlating well with experimental results suggesting limited electron doping in monolayer Re-doped $WSe_2$[27]. On the other hand, some dopants (e.g. V) may be less strongly-localized, even at the monolayer limit, such that they can modify transport, enabling true monolayer junction engineering[31]. Further understanding of the impact of dopant localization on photonic and electronic properties is required to fully access the potential of monolayer and few-layer TMDs.

## The Physics of Defects in 2D Semiconductors

In Figure 4 we show differential conductance spectra for $V_W^-$, $Cr_W$, $Mo_W$ and $Re_W^+$ in monolayer $WSe_2$. As expected, vanadium introduces a series of defect states above the valence band maximum, while for rhenium we find defect states below the conduction band minimum. Because the TMD is grown on epitaxial graphene on SiC that pins the Fermi level ($V$ = 0 V in the dI/dV spectra), the dopants are ionized, adopting a negative charge state for V and a positive charge state for Re. This leads to a distinct appearance in the STM topography where a negatively (positively) charged defect appears as extended depression (protrusion) at positive and an extended protrusion (depression) at negative sample bias. In addition to their chemical defect state(s), an ionized defect may also give rise to a series of hydrogenic states. These are bound states emerging from the screened Coulomb potential localized at the charged defect[35]. Problematically, in 2D TMDs they have a similar spatial extent like chemical states such that they cannot be easily discriminated from each other without theoretical modelling.

Generally, a defect's electronic[25,26,35], optical[36] and magnetic spectrum[28] is determined by the electron configuration of the impurity, the coordination to the host lattice, charge-transfer from/to the substrate, and screening by the substrate. Moreover, spin-orbit coupling is a highly relevant interaction in these transition metal host crystal and compounds that leads to considerable defect state splitting as observed for the sulfur vacancy ($Vac_S$[26]) and $Cr_W$[25] in $WS_2$. Furthermore, point defects can strongly couple to phonons, which we analyzed in detail for a single carbon radical ion (CRI) in $WS_2$[28]. The CRI exhibits a significant electron-phonon coupling, with a Huang Rhys factor of up to 5.4, for a dominant local vibrational mode. Interestingly, the coupling strength depends strongly on the charge and spin state and decreases with increasing number of layers. While these results are specific to this particular type of defect, they highlight the general importance of electron-phonon coupling for deep defect states with tightly confined wavefunctions commonly observed in 2D semiconductors.

While we summarized here some recent advancements in engineering and probing atomic defects/dopants in 2D TMDs, many challenges remain regarding modeling, synthesis, characterization, manipulation and device integration. In the following we outline some open challenges that may serve as a guideline for future work in this area.

## Future Prospects for Probing and Engineering Quantum Defects in TMDs

While not covered in the preceding discussion, **theoretical modelling** of defect systems in TMDs have greatly advanced our understanding of the complex physics of defects in low-dimensional materials systems. In particular the interlinking of electronic structure theory and many-body optics have been a great breakthrough[37]. The increasing predictive power of *ab initio* modelling of ground state properties will guide structure exploration more strongly in the future. Modelling the dynamic propagation and coherence of a system in a complex environment represent the next level in these efforts.

Despite many appealing reasons to use semiconducting TMDs in next-generation electronics, industry adoption will largely depend on the ability to **reproducibly synthesize** doped 2D semiconductors in sufficient quality at growth conditions compatible with existing process standards. First steps in this direction are currently taken, achieving wafer-scale doping uniformity at front- and back-end-of-line compatible temperatures[27,33,32,31]. Standardized benchmarking of the material quality and doping density/distribution will be critical to monitor the process quality. Elemental and isotopic purification steps at the precursor level[38] or post-synthetic chemical surface treatment will be necessary to remove even low levels of impurities that can have detrimental effects on charge carrier dynamics or optical properties[39,40].

To explore the rich atomic, electronic, spin and optical properties of defects in 2D materials, more **advanced scanning probe methods** are currently developed that push the space, energy and time resolution to their ultimate limits. Nc-AFM with functionalized tips (Figure 5a) enables to unambiguously assign the defect site and identify minute details in the local strain profile[25]. Near-field scanning probe techniques can characterize light-matter interactions such as single-photon emission with nanometer scale precision. Using electrically-driven photon emission from an STM (Figure 5b) it was even possible to map the spatial light distribution emitted by a single defect[36]. Accessing the magnetic properties of defects and manipulate their spin will be another important frontier in the field. It has been demonstrated that for certain point defects the electron spin can be addressed optically *via* optically-detected magnetic resonance[41,42]. The recently introduced electron spin resonance STM (ESR-STM) technique[43] using an external magnetic field, spin-polarized tips and radio frequency electric fields to resonantly drive spin transitions maybe uniquely suited to probe the spin manifold of single defects at µeV energy resolution (Figure 5c). This method can also resolve the nuclear spin *via* the hyperfine interaction[44]. Moreover, pulsed ESR-STM can be used to coherently manipulate the electron spin on the Bloch sphere[45]. Ultrafast STM is another emerging frontier, which is expected to yield significant insights into the excitation dynamics at single defects. The recent advances in lightwave-driven tunneling using phase-stable THz pulses coupled to an STM (THz-STM, Figure 5d) introduced by Cocker *et al.*[46,47], demonstrated femtosecond time resolution at atomic spatial resolution. Both the ESR-STM and THz-STM techniques offer new exciting possibilities to probe the time dynamics of charge, spin and optical excitations and understand their decoherence channels.

On an ***engineering*** level, we anticipate three major challenges: (i) the deterministic lateral placement of defects, (ii) gate-tunable charge and spin state control, and (iii) developing an integrated optical interface between defects for their controlled coupling. Especially the first point is a formidable task that needs to be addressed. Periodic large-scale defect patterns may be established by synthetic self-organization. Domain boundaries could serve as a predefined docking site for dopants forming a 1D channel (Figure 5e). Alternatively assemblies of precise nanoplatelets or large TMD precursors that contain a dopant could form a regular 2D pattern (Figure 5f). Top-down approaches are limited to methods with the required sub-nanometer placement control, such as scanning transmission electron microscopy (STEM, Figure 5g) manipulation, Helium focused ion beam (FIB, Figure 5g), or STM atomic manipulation (Figure 5h,i). Each of these techniques has its limitations when it comes to TMDs: repositioning atoms with an electron beam demonstrated for impurities in graphene[48] may not be generally applicable to different 2D platforms such as TMDs given their distinct bond strengths and conductivity. Moreover, the large kinetic energy of the electron beam may also cause unwanted damage to the material. Focused He ion beams were shown to create single point defects such as chalcogen vacancies with a lateral precision below 10 nm on a substrate[29]. While this precision may be sufficient for optical applications that are diffraction limited, it is insufficient for creating spin structures where interaction varies on a sub-angstrom length scale. Despite its first demonstration dating back to 1990, the capability of STM atomic manipulation to write complex patterns with single adatoms on a surface continues to impress[49]. Lateral manipulation relies on a weak binding energy between adatoms and the surface, a condition that is usually not fulfilled for a covalently bonded impurity. However, adatoms may be moved to the desired location first, where they get activated for instance by a voltage pulse to pin them on the surface or drive them into the material (Figure 5h). This hypothetical concept would combine the flexibility of atomic manipulation without sacrificing the stability of a covalently bonded impurity. Alternatively, one could chemically treat the surface to introduce weak, well-defined breaking points. Applying a voltage pulse at specific sites would then expose a dangling bond that could be used as an anchoring point for subsequently deposited atoms (Figure 5i). A first proof-of-principle of this concept has been demonstrated by STM-induced hydrogen desorption at $CH_S$ defects in $WS_2$ introduced by a post-synthetic methane plasma treatment[28].

Despite these challenges ahead, 2D semiconductors offer exciting possibilities to engineer atomic quantum systems by chemical design principles from the bottom up. New synthetic routes, advanced characterization methods and ingenious device concepts are continuing to unveil the true potential of atomically-engineered 2D materials platforms.

## Acknowledgements

J.A.R. acknowledges Intel through the Semiconductor Research Corporation (SRC) Task 2746, NSF-DMR 2002651, and the Penn State 2D Crystal Consortium (2DCC)-Materials Innovation Platform (2DCC-MIP) under NSF cooperative agreement DMR-1539916 for financial support. B.S. appreciates funding from the European Research Council (ERC) under the European Union's Horizon 2020 research and innovation program (Grant agreement No. 948243).

## Conflict of Interest

Authors declare no conflict of interest.

## Data Availability Statement

The data that support the findings of this study are available from the corresponding author upon reasonable request.

# Figures

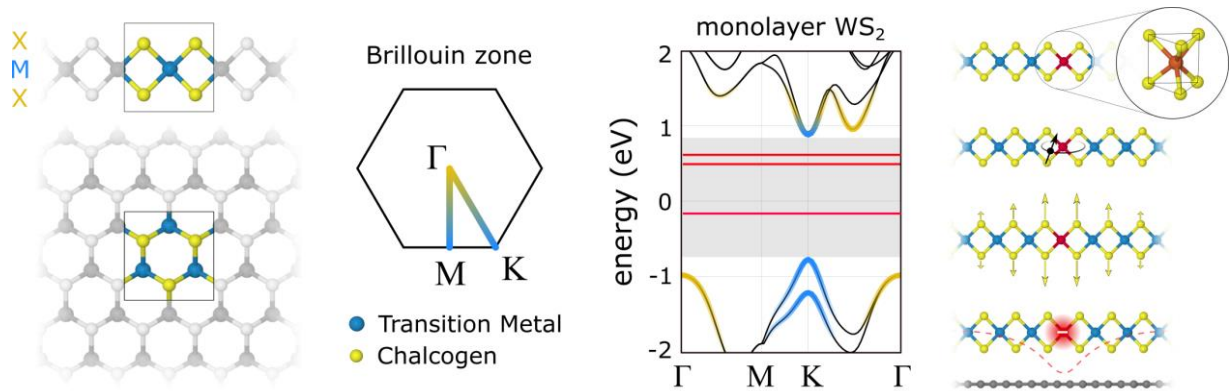

*Figure 1: **Semiconducting Transition Metal Dichalcogenide Monolayers**. Discrete electronic states (red lines) are introduced in the TMD band gap. The energetic defect spectra depends on the electron configuration of the impurity, the coordination to the host lattice, charge-transfer from the substrate, and screening by the substrate.*

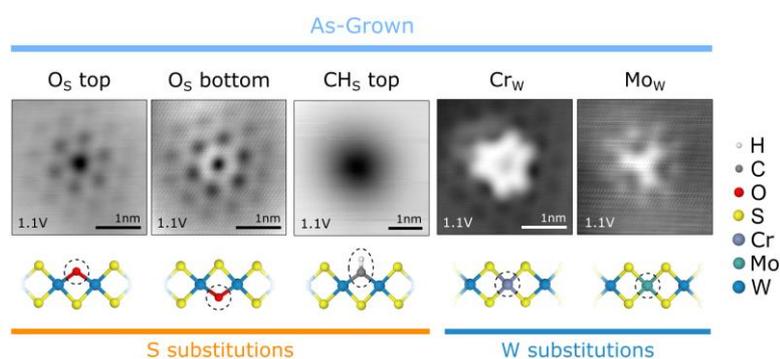

*Figure 2: Common **accidental point defects** in as-grown CVD TMDs [25]. Reprinted (adapted) with permission from Schuler et al. ACS Nano 12, 10520 (2019). Copyright 2021 American Chemical Society.*

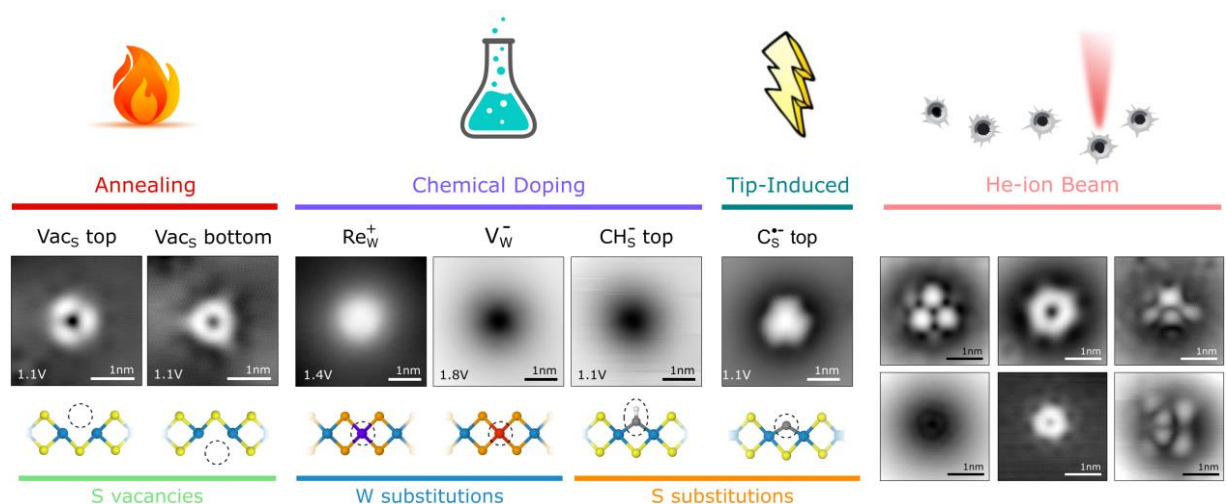

*Figure 3: **Intentional defects** and dopants induced by in-vacuo annealing, chemical doping, atom manipulation and ion beam bombardment.*

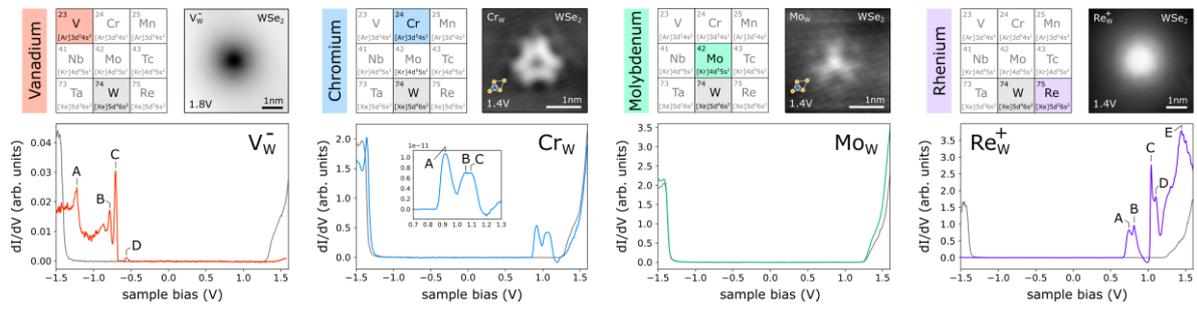

Figure 4: Electronic spectra of **transition metal substitutes**: $V_W^-$, $Cr_W$, $Mo_W$ and $Re_W^+$ [27] in monolayer $WSe_2$.

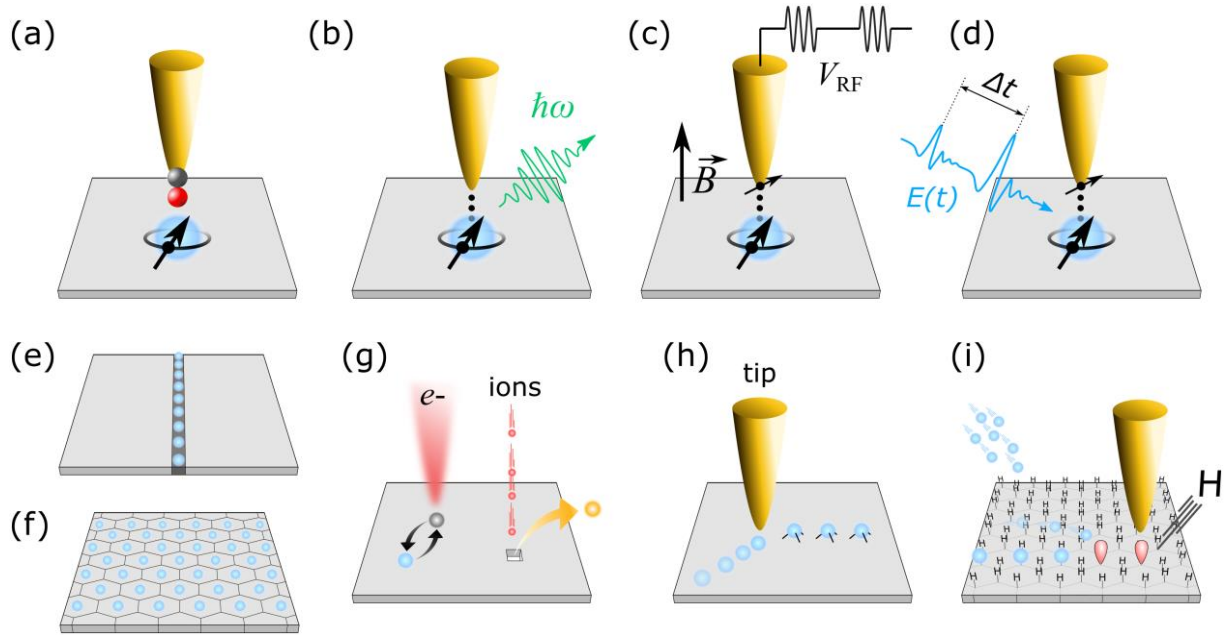

Figure 5: (a-d) **Advanced scanning probes for ultimate resolution in space, energy and time**. (a) Tip functionalization (e.g. CO) to increase lateral resolution. (b) STM luminescence to investigate light-matter interaction at atomic scale. (c) ESR-STM with spin-polarized tips to probe spin manifold with µeV energy resolution. (d) Pump-probe THz-STM to probe the time dynamics of the excitation spectrum. (e-i) Possible concepts for **lateral placement control** of point defects (blue spheres). (e,f) Synthetic self-organization, e.g. along domain boundaries (e) or using well-defined nanoplatelets (f). (g) Atomic manipulation with electron (left) or ion beams (right). (h) Atomic manipulation by a scanning probe tip to move surface atoms/molecules and fix/implant them into the host matrix. (i) Tip-induced desorption of chemically-treated 2D materials exposing dangling bonds (red) as anchoring points for dopants.